\documentstyle[prb,aps]{revtex}
\input epsf.sty
\begin{document}
\draft
\title{Monte Carlo Study of Short-Range Order and Displacement Effects in 
Disordered CuAu}
\author{O. Malis, K. F. Ludwig, Jr.}
\address{Boston University, Boston, MA.}
\author{D. L. Olmsted, B. Chakraborty}
\address{Brandeis University, Waltham, MA.}
\date{\today}
\maketitle
\begin{abstract}

The correlation between local chemical environment and atomic
displacements in disordered CuAu alloy has been studied using Monte
Carlo simulations based on the effective medium theory (EMT) of metallic
cohesion.  These simulations correctly reproduce the chemically-specific
nearest-neighbor distances in the random alloy across the entire
Cu$_x$Au$_{1-x}$ concentration range.  In the
random equiatomic CuAu alloy, the chemically specific pair distances
depend strongly on the local atomic environment (i.e. fraction of
like/unlike nearest neighbors).  In CuAu alloy with short-range order,
the relationship between local environment and displacements remains
qualitatively similar.  However the increase in short-range order causes
the average Cu-Au distance to decrease below the average Cu-Cu distance,
as it does in the ordered CuAuI phase.  Many of these trends can be
understood qualitatively from the different neutral sphere radii and
compressibilities of the Cu and Au atoms.

\end{abstract}

\section*{1. Introduction}

Understanding the detailed structure of disordered metallic alloys continues to 
be an experimental and theoretical challenge.  Particularly interesting
are alloys in which there is a significant mismatch in atomic size.
Intuitively, it might be expected that the distances between nearest neighbor
pairs in an alloy with large atoms (A) and small atoms (B) would follow 
a
simple  relationship $r_{AA} > r_{AB} 
> r_{BB}$.  However, x-ray and neutron diffuse scattering experiments
indicate that the behavior is more complex and that there are correlations
between chemical ordering tendencies and pair distances (Jiang, Ice, Sparks, Robertson and Zschack 1996). 
Unfortunately, our general understanding 
is limited by the statistical nature of the scattering process -- 
typically only average pair distances and chemical coordination numbers can be 
determined 
by these methods.  They do not provide information about the relationship 
between local displacements from ideal lattice sites and the local degree 
of short-range order.
Using an effective medium theory (EMT) energy calculation
(Jacobsen, Norskov and Puska 1987), we
present here a detailed Monte Carlo (MC) study of the correlation between 
local order and displacements in
equiatomic disordered CuAu (12\% size difference).   EMT is a semi-empirical
approach belonging to the category of embedded atom methods, and is based on
Density Functional Theory (Jacobsen et al. 1987) .
These simulations offer the advantage
of giving access to individual atoms and their local chemical environment,
thus providing information complementary to experiment.
They exhibit clear correlations between local chemical environment and 
displacements in disordered CuAu alloy.

At 50-50 at.\% concentration CuAu has two first order phase transitions.
Above 683K the stable phase is a FCC disordered phase. Below 658K the stable 
phase CuAuI is an ordered $L1_0$ phase with a 7\% tetragonal distortion.
Between those two temperatures the stable phase is a long-period superlattice,
CuAuII, with a modulation wavevector perpendicular to the ordering wavevector. 
The wavelength of the periodic anti-phase boundaries is ten times 
the size of the underlying ordered cell.

\section*{2. Monte Carlo Simulations Based on Effective Medium Theory}

\subsection*{2.1. Simulation Model Details}

 First principles calculations of the effects of atomic displacements on alloy
phase stability have often been based on effective Ising models where the positional
degrees of freedom are integrated out at zero temperature
(Lu, Laks, Wei and Zunger 1994, Wolverton and Zunger 1995). 
Finite temperature simulations involving both configurational and positional
variables have been possible only with computationally efficient empirical 
potentials (Dunweg and Landau 1993, Polatoglou and Bleris 1994, Silverman, 
Zunger, Kalish and Adler 1995).
The advantage of a direct simulation over simulations based on effective Ising
models is that they yield detailed information about the nature of atomic
relaxations and the coupling between configurational and positional degrees of
freedom: information which is hidden in higher order correlation functions of
effective Ising models. 
These considerations led us to investigate the role of atomic displacements in
Cu-Au alloys using the EMT formalism. Though EMT is not a first-principles 
approach,
it is a useful model which captures some essential features of
metallic binding.

A number of researchers have found that EMT can reproduce 
well the bulk and surface properties of many pure metals (Jacobsen 1988).
EMT-based calculations have also been successfully applied 
to investigate equilibrium and kinetics properties of Cu-Au alloys.
Monte Carlo simulations using the EMT have correctly reproduced the 
order-disorder transitions in the Cu-Au phase diagram
(Xi, Chakraborty, Jacobsen and Norskov 1992). This work 
illustrated the importance of going beyond cubic fixed-lattice models
to include tetragonal distortions (Xi et al. 1992) and atomic
displacements (Chakraborty 1995).  In addition, EMT has been used to construct 
an atomistic Landau theory of the alloy (Chakraborty and Xi 1992)
which was able to predict qualitatively the stability
of the modulated phase CuAuII in a narrow temperature range. Langevin 
simulations
of the ordering kinetics using the Landau model were 
recently found to be in qualitative agreement with experimental x-ray 
results (Elder, Malis, Ludwig, Chakraborty and Goldenfeld 1998).

The EMT formalism provides a structure for systematically constructing 
interatomic potentials in metallic systems where simple pair potentials are
known to be inadequate.  These potentials belong to the same category as
embedded atom potentials (Daw, Foiles and Baskes 1993), however, the details 
of the construction are different (Jacobsen et al. 1987). The EMT interactions 
involve parameters characterizing
atoms in specific environments; for example, the spatial extent of the
electron density distribution around a Cu atom 
embedded in an electron gas of given density (Jacobsen et al. 1987). In
principle, these parameters can be obtained from {\it ab initio} calculations
based on DFT. However, in its application to the Cu-Au alloys
(Xi et al. 1992), a semi-empirical
approach has been adopted where the atomic parameters are obtained from fitting
to the ground-state properties of the {\it pure} metals and an additional
parameter, which enters only in the description of an alloy, is obtained by
fitting to the formation energy of CuAu (Stoltze 1997).

We performed MC simulations of CuAu in the canonical ensemble using the EMT 
approach.
In our MC implementation the Metropolis algorithm is used to determine
the acceptance/rejection of 3 different kinds of system changes:
the interchange of Cu and Au atoms, the displacement of individual atoms
from their ``ideal'' lattice sites, and the size of the global lattice 
constants.
The candidate atomic displacements are chosen randomly in a box whose
size is adjusted to optimize the MC acceptance rate.  
The atomic displacements are due both to thermal vibrations and to ``size 
effects''.

The simulations are performed using a modified version of the ARTwork simulation 
package (Stoltze 1997) running on a Silicon Graphics Origin2000 computer system.
 One full MC step for an $N$-atom simulation cell ($32^3$ or $60^3$) 
consists of $N$ attempts to (a) randomly exchange atoms, and (b)  change 
the position of the atoms involved in the exchange followed by 
one attempted change in global lattice constants. When examining properties 
of the disordered phase the cubic symmetry is fixed. Typically, the alloy is 
equilibrated at temperature for 1000 MC steps before the data is stored 
for processing. The averages are usually taken over 50 configurations 
saved every 10 MC steps. 

\subsection*{2.2. Model Accuracy}

Before discussing correlations between local chemical order and atomic 
displacements in our Monte-Carlo simulations of disordered CuAu, we examine 
the accuracy of the EMT model. The model predicts the correct Cu$_3$Au, CuAu
and Au$_3$Cu regions of the phase diagram. 
For the 50-50 composition studied in detail 
here, previously published EMT simulations which did not allow atomic 
displacements found (Xi et al. 1992) that
the model presents a first order phase transition at
approximately 708K between the cubic disordered phase and the tetragonal 
ordered phase.  The ratio of the lattice constants in the ordered phase $c/a$ is 
0.94 and agrees well with the actual value of 0.93.  In the new simulations 
reported here, atomic displacements are allowed, but the transition remains 
first order in accord with experiment. The jump of the long-range order
parameter at the phase boundary is 0.94.  In addition, the ratio of the 
lattice constants $c/a$ in the ordered phase
remains 0.94.  However the transition temperature is considerably lowered
by the inclusion of atomic displacements to approximately 430K.  Since
displacements due to both thermal vibrations and the ``size effect''
would be expected to preferentially lower the free-energy of the
disordered phase, the drop in transition temperature is not surprising. 
Thus the inclusion of atomic displacements decreases
the agreement with the experimental transition temperatures when
using the parameters of Xi et al. 1992.
As mentioned above the alloy parameter entering the EMT model was obtained
by fitting to the CuAu formation energy without incorporating displacements. 
Therefore it would be possible to improve the
current model by adjusting the EMT parameters.  However, 
for consistency with the previously published simulations, we chose to keep 
the parameters from Xi et al. and to scale all temperatures 
with respect to the new transition temperature.  

\begin{figure}[tbh]
\begin{center}
\epsfysize=10cm
\epsfbox{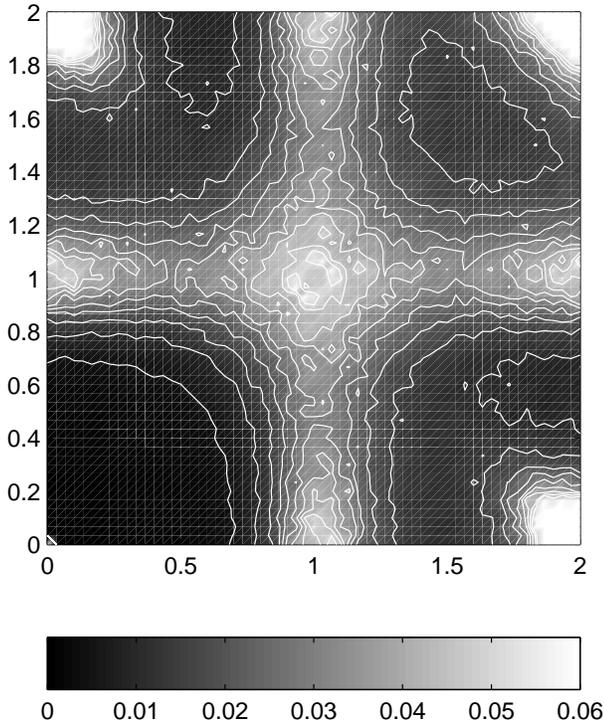}
\end{center}
\caption{Diffuse scattering intensity in the (hk0) reciprocal plane from a 
$60^3$ atom simulation of CuAu 2.7\% above the transition temperature. 
The intensity is normalized to $N^2 x_{Cu} x_{Au}$. $f_{Cu}$ and $f_{Au}$ were 
considered constant and taken to be 26 and 72, respectively. }
\label{diffuse}
\end{figure}

The local chemical order and atomic displacements in disordered alloys are often 
measured by diffuse x-ray or neutron scattering. For comparison, we calculated 
the x-ray diffuse scattering intensity for the EMT model by Fourier 
transformation for several effective temperatures just above the phase 
transition.  A typical diffuse scattering in the $(hk0)$ reciprocal plane 
calculated from 
a simulation 2.7\% above the transition temperature is presented in 
Fig.\ref{diffuse}. The asymmetry of the diffuse scattering around the
superlattice points is due to atomic displacements (Borie and Sparks 1971).
Though not easy to see on the scale of the plot, the 
simulated peaks exhibit the anisotropic
four-fold splitting observed in experiment (Hashimoto 1983, Malis, Ludwig, 
Schweika, Ice and Sparks 1998). This splitting, 
which is due to correlations extending well beyond a few unit cells, is 
reminiscent of that produced by the long-period superlattice of CuAuII.  
Simulations at several temperatures in the disordered phase reveal that the 
superlattice peaks grow relatively slowly with decreasing temperature, in 
accord with experimental observation (Malis et al. 1998).

Metcalfe and Leake (Metcalfe and Leake 1975) have reported a study of the 
local chemical order in CuAu alloy as measured by x-ray diffuse scattering.  
They analyzed 
their experiments in terms of short-range order parameters $\alpha_{lmn}$ which 
measure the average number of like/unlike neighbors surrounding an atom in a 
given shell (Borie and Sparks 1971):

\begin{equation}
\alpha_{lmn} = 1 - {{P_{AB}^{lmn}}\over{x_B}}
\end{equation}

\noindent Here $P_{AB}^{lmn}$ is the probability of having a B atom at the 
$(lmn)$ lattice site if there is an A atom at the origin.
In a completely random alloy $\alpha$ is zero. Positive values of $\alpha$ 
indicate that the corresponding shell is mainly occupied by atoms of the same 
kind as the atom at the origin. If $\alpha$ is negative the number of unlike 
neighbors is dominant.  For comparison we have calculated the simulated
$\alpha_{lmn}$ directly in real space at an effective temperature 2.7\% above 
the 
phase transition.  The calculated values are compared with those from 
Metcalfe and 
Leake in Table \ref{talphas}.  The simulated values of the SRO parameters for 
the first two 
coordination shells are 2.5\% and 10\% lower than the values reported by
Metcalfe and Leake for a sample at a comparable temperature (700 K).  Given 
uncertainties in the accuracy of the experimental $\alpha$'s 
\footnote{Metcalfe and Leake used a very approximate 
scheme to eliminate the thermal diffuse scattering, which is quite large. 
Moreover, they ignored second-order static displacements, which can contribute 
significantly to the scattering in their wavevector range.  Direct evidence of 
the limited accuracy of their results is that their $\alpha_{000}$ values, 
which should by definition be unity, are 26-39\% too high.} 
this close agreement may well be fortuitous.
The simulations also correctly predict that the magnitude of the third neighbor 
correlations drops by approximately an order of magnitude compared to those for 
the two nearest shells.  However, the small remaining correlation is incorrectly 
predicted to be positive for this shell.  As a result, the tails of the 
simulated diffuse scattering peaks shown in Fig. \ref{diffuse} are less 
symmetric than is the actual case in the alloy.  Nonetheless, 
we can conclude that the degree of local order predicted by the EMT Monte-Carlo 
model is quite reasonable.

\begin{figure}[tbh]
\begin{center}
\epsfysize=10cm
\epsfbox{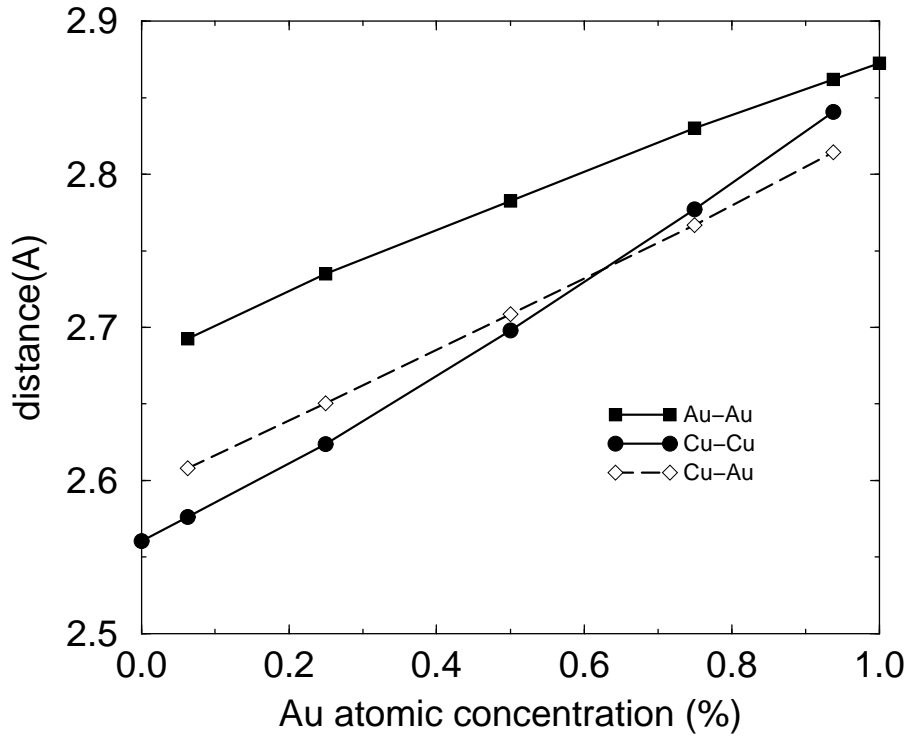}
\end{center}
\caption{Concentration dependence of the average Cu-Cu, Au-Au and Cu-Au
distances in a random alloy at low temperature. The system size for these 
simulations was 2048 atoms.
}
\label{random}
\end{figure}

Frenkel et al. (Frenkel, Stern, Rubshtein, Voronel and 
Rosenberg 1997) have used EXAFS to measure the average nearest 
neighbor 
distances in Cu$_x$Au$_{1-x}$ alloys with random atomic arrangements at 80 K.  
To better test the accuracy of the EMT
Monte-Carlo simulations, we performed simulations of a random CuAu 
alloy quenched to an equivalent temperature and determined the average 
nearest-neighbor distances for the Cu-Cu, Cu-Au and Au-Au pairs. These are shown in 
Fig. \ref{random}.
At all compositions the simulated distances agree with the EXAFS measurements
within the experimental uncertainty.  Interestingly, both the EXAFS data and our
simulations show that the average Cu-Au distance is very close to the Cu-Cu 
distance 
and significantly less than the Au-Au distance.  Moreover, our simulations 
predict
a crossover of the Cu-Cu and Cu-Au distances near 60\% Au concentration, with
the Cu-Au distance being the smallest pair distance above this concentration.
This is in agreement with first-principles calculations of 
Ozolins et al. (Ozolins, Wolverton and Zunger 1997). 
It is noteworthy that the strong concentration
dependence of the relative Cu-Cu and Cu-Au distances is larger than expected
from compressible Ising models incorporating simple elastic energy and 
displacement-spin coupling terms (Chakraborty 1995).

The behavior of nearest-neighbor distances, shown in Figure \ref{random}, can 
be 
rationalized on the basis of the EMT description of cohesion.  In this
description, the Au atoms have a larger neutral-sphere radius (the ``size'' in
EMT) than the Cu atoms (Jacobsen et al. 1987).  Moreover, Au is less compressible.  In
the simplest approximation, we can view the increase in nearest-neighbor distances
with increasing Au concentration as a traditional steric effect associated with
the larger EMT ``size'' of Au atoms.  
Though this trend is definitely observed in Figure \ref{random}, the rate of
increase is different for the three different types of chemical bonds.  This can 
be attributed to the difference in compressibilities of Cu (lower bulk modulus -- $B_{Cu}$ =
14 x 10$^{10}$ Pa) and Au (higher -- $B_{Au}$ = 17 x 10$^{10}$ Pa).
Although there is some bowing present in the Cu-Cu line, it largely
follows the changing lattice parameter with increasing Au content (note
that the Cu-Cu pair distance extrapolates to the Au-Au distance at the
highest Au concentrations -- i.e. to the pair distance in pure Au).
Thus the relatively higher compressibility of Cu atoms allows the Cu-Cu
distances to follow the ``average'' lattice with changing concentration.
The relative incompressibility of the Au atoms, however, prevents the
average Au-Au distance , and the Cu-Au distance, from changing as much with 
concentration.  Within the EMT model, the relative compressibilities can also 
be related to the more rapid falloff of the electron density around a Au atom
than around a Cu atom at typical interatomic distances.

\section*{3. Correlations between local environment and displacements}

\begin{figure}[tbh]
\begin{center}
\epsfysize=14cm
\epsfbox{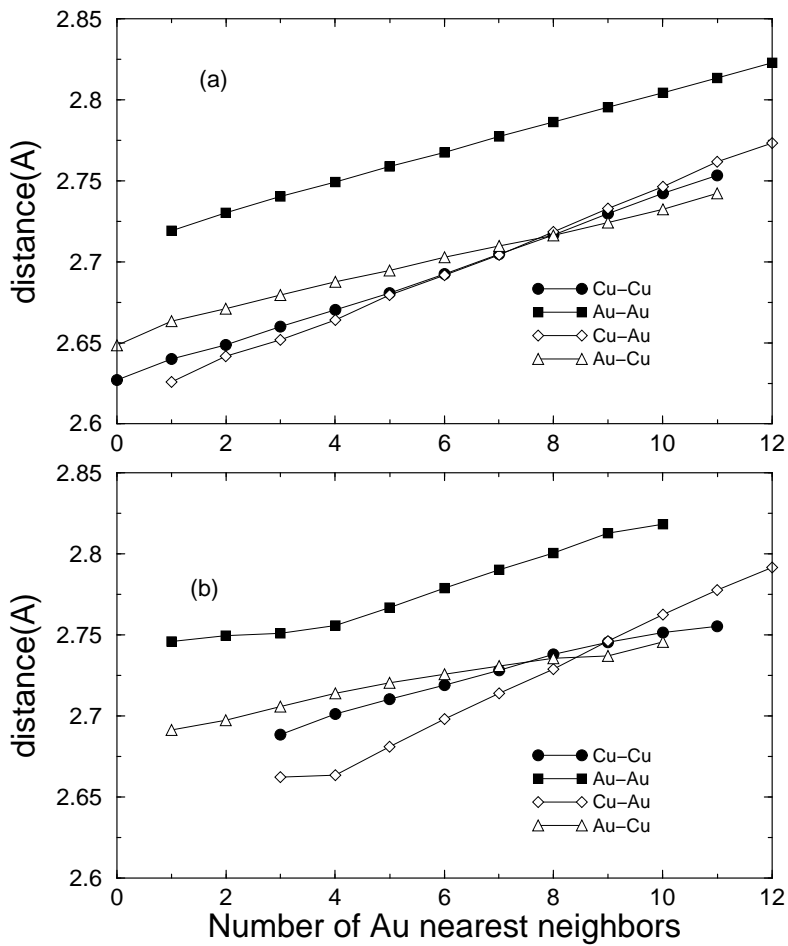}
\end{center}
\caption{Dependence of the Cu-Cu, Au-Au, Cu-Au (Cu central atom) and Au-Cu 
(Au central atom) nearest-neighbor distances 
on the number of Au nearest neighbors in (a) a random equiatomic CuAu alloy 
at low temperature and (b) in the same alloy with short range order 2.7\% above 
the transition 
temperature.}
\label{auconc}
\end{figure}

Though random alloys exhibit no global short-range order, the
constituent atoms nonetheless experience stochastic variations in their
nearest-neighbor environments.  Therefore, in order to investigate the
relationship between local environment and pair distances, we have
calculated the average interatomic distances in the random equiatomic
CuAu alloy as a function of the number of nearest-neighbor Au atoms.
That is, we have divided the atoms in the random CuAu alloy into 12
groups -- each member of the group has the same number of Au
nearest-neighbors.  We have then calculated the average Cu-Cu, Cu-Au (Cu
being the central atom), Au-Cu (Au being the central atom) and Au-Au
pair distances in each group.  The results are shown in Figure \ref{auconc}(a).  
There is a strong relationship between local environment and pair
distances.  The trends with increasing Au in the local environment are
qualitatively similar to those seen in Figure \ref{random} with increasing 
overall Au concentration.  Clearly, even in a random alloy at a
fixed concentration, the intuitive idea of atomic ``size'' has limited
meaning.  

\begin{table}
\caption{Comparison
of the short-range order parameters $\alpha_{hkl}$ obtained 
from a simulation 2.7\% above the transition temperature with the experimental 
values (Metcalfe and Leake 1975) for a quench from 700K (2.5\% above the 
true transition temperature). The last column shows the values for 
perfectly ordered CuAu.}
\begin{tabular}{dddd}
hkl&simulation&experiment&ordered CuAu\\
\tableline
000&1.0000&1.263&1.00\\
110&$-$0.1825&$-$0.187&$-$0.33\\
200&0.2069&0.230&1.00\\
211&0.0226&$-$0.013&$-$0.33\\
220&0.0484&0.109&1.00\\
310&$-$0.0557&$-$0.029&$-$0.33\\
\end{tabular}
\label{talphas}
\end{table}

\begin{table}
\caption{Simulated nearest-neighbor relative displacements
$r_{AB} = (d_{AB}-d_0)/d_0$,
where $d_0$ is the average distance, at a temperature 2.7\%
above the transition temperature and infinite temperature (random alloy),
respectively.}
\label{tdist}
\begin{tabular}{ddd}
pair&finite temperature&random configuration\\
\tableline
$\alpha_{110}$&$-$0.182&$\approx$ 0\\
Cu-Cu&0.0015&$-$0.0092\\
Au-Au&0.0171&0.0223\\
Cu-Au&$-$0.0016&$-$0.0051\\
\end{tabular}
\end{table}

In order to better understand the relationship between local environment
and displacements in a non-random alloy, we studied in detail a
simulated equiatomic CuAu alloy annealed at a temperature 2.7\% above the
ordering transition to produce short-range order.  In direct contrast to
the case for random equiatomic alloys, the overall average Cu-Au distance is
shorter than the overall average Cu-Cu distance (table \ref{tdist}).  Thus the
development of short-range order has changed the average pair
distances.   Figure \ref{auconc}(b) shows how the average Cu-Cu, Cu-Au, Au-Cu 
and
Au-Au pair distances vary with number of Au nearest-neighbors.
Comparison with the equivalent Figure \ref{auconc}(a) for a random alloy shows 
that the overall trends remain qualitatively similar but the short-range order
causes quantitative changes.

\begin{figure}[tbh]
\epsfysize=14cm
\epsfbox{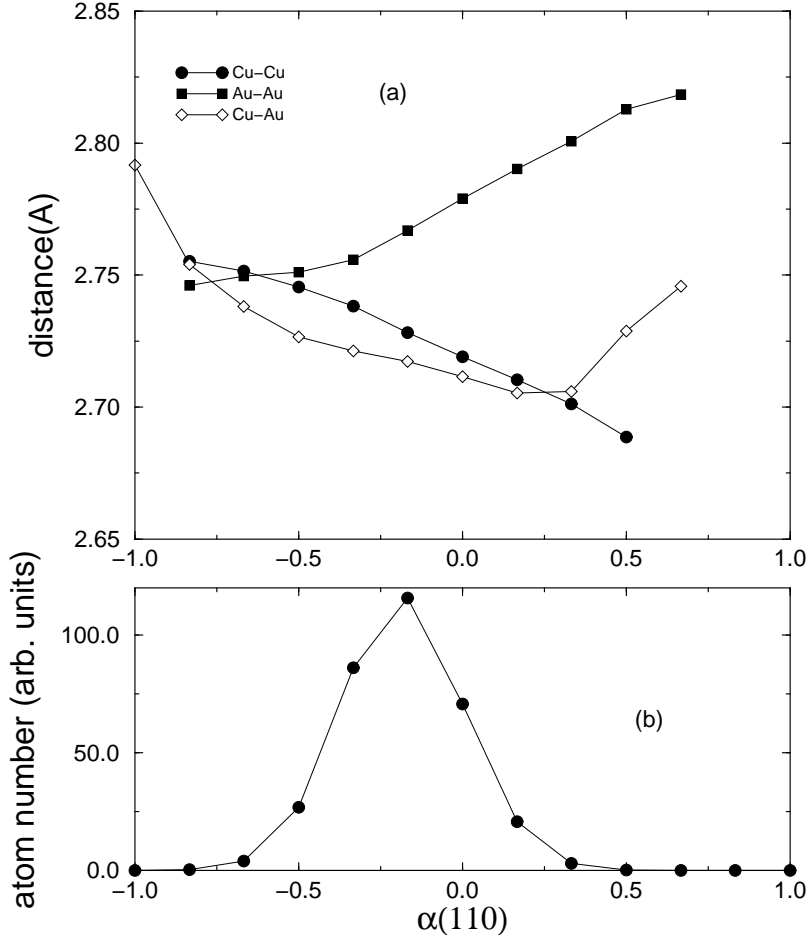}
\caption{(a) Dependence of the Cu-Cu, Au-Au and Cu-Au nearest-neighbor distances 
on the local coordination number $\alpha_{110}$ 2.7\% above the transition 
temperature. The error bars are smaller than the size of the symbols.
(b) The distribution of local $\alpha_{110}$ values.}
\label{scatter}
\end{figure}

Another way of examining the data from the alloy with short-range order
is to define a local $\alpha_{110}$ for each atom -- i.e. the number of
like nearest neighbors minus the number of unlike nearest neighbors
divided by the coordination number.  Figure \ref{scatter}(a) shows the 
systematic
relationship between the local ``order'' as measured by this
$\alpha_{110}$ and the interatomic distances.  Here we have grouped
together the Cu-Au and Au-Cu distances.  Figure \ref{scatter}(b) shows also the
histogram of local $\alpha_{110}$ values.  For atoms with large positive
values of $\alpha_{110}$, i.e. a large fraction of like nearest
neighbors, the relation between like and unlike pair distances follows
the simple expectation that $d_{Cu-Cu} < d_{Cu-Au} < d_{Au-Au}$.
However, for lower and negative values of $\alpha$ the Cu-Au distance is
the smallest of the three.  Moreover, the Au-Au distance decreases and
begins to approach the Cu-Cu distance, which itself increases with
decreasing $\alpha$.  The behavior is very suggestive of the
relationships between nearest neighbor distances in the ordered CuAuI
phase - i.e. $d_{Cu-Cu} = d_{Au-Au} > d_{Au-Cu}$.  It can also be seen
from Figure \ref{scatter} that the development of short-range order, i.e. 
an increase in the number of atoms with a negative $\alpha$, in the alloy leads 
to a decreasing average Cu-Au distance relative to the average Cu-Cu
distance.  This partly explains why $d_{Cu-Au} < d_{Cu-Cu}$ in the
equiatomic alloy with short-range order but $d_{Cu-Au} > d_{Cu-Cu}$ in
the random alloy.

\begin{figure}[tbh]
\begin{center}
\epsfysize=10cm
\epsfbox{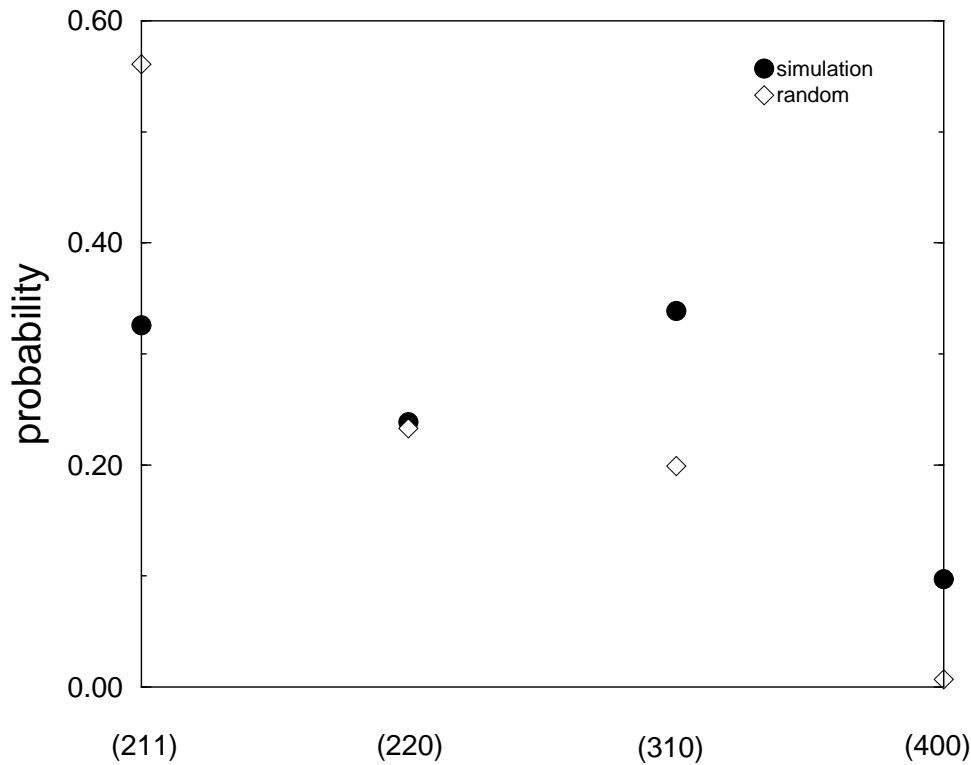}
\end{center}
\caption{Probabilities of finding the 4 like nearest neighbors
of those atoms with $\alpha_{110} = -0.33$ distributed in the 3 surrounding 
planes
at the simulated temperature. Also shown are the random probabilities.
}
\label{planes}
\end{figure}

We studied closely the atoms positioned in a local chemical environment similar
to the environment in the ordered phase, i.e. having 4 like and 8 unlike nearest 
neighbors (i.e. local $\alpha_{110} = -0.33$). In the ordered phase the 4 like 
neighbors occupy one of the cubic planes passing through the center atom while 
the 8 unlike atoms occupy the other 2 planes. In the simulation we found that 
approximately 10\% of the atoms with 8 unlike nearest neighbors are actually in 
a geometrically ``ordered'' environment. However,
the occurrence of this configuration is considerably higher than random. 
Figure \ref{planes} contrasts the occupation probabilities in the planes for the 
random case and that observed in the EMT simulations.  It is instructive to 
focus on those atoms in a geometrically ``ordered'' environment (as well as 
chemically ordered environment) and define a local distortion as the 
deviation from unity of the ratio of the average A-A distance to the average A-B 
distance around that atom.  The average local distortion around Cu 
atoms in the simulation is approximately 1.5\%, slightly higher than the average 
around Au atoms (0.4\%). Both values are much smaller than the 3\% that would be 
predicted by the tetragonal distortion occurring in the ordered structure.

\section*{4. Conclusions}

An important advantage of real space simulations is that they permit a
direct examination of the relationship between local chemical
environment and atomic displacements.  These simulations show a strong
correlation between nearest-neighbor environment and interatomic
distances in disordered Cu$_x$Au$_{1-x}$, even in random alloys.  Higher
Au concentrations in the local environment lead to a decrease of Cu-Au
distances relative to Cu-Cu distances.  In equiatomic CuAu alloys with
short-range order, these trends remain qualitatively similar.  The
increased local ``order'' causes a decrease in the average
nearest-neighbor Cu-Au distance and a partial convergence of Cu-Cu and
Au-Au distances, reminiscent of the local structure in the ordered CuAuI
phase.  The intuitive concept of ``size'' has limited  usefulness in
understanding these trends.  We anticipate that such correlations
between local environment and displacements are common in disordered
alloys, particularly those having components with large atomic size
differences.

\section*{Acknowledgments}

We would like to acknowledge useful discussions with C. J. Sparks and W. 
Schweika.
This work utilizes the resources of the Boston Univ. Center for Computational
Science and was supported by NSF grant DMR-9633596.
B. C. would like to acknowledge numerous discussions with Per Stoltze regarding
the ARTwork program.  The work of B. C. and D. O. was supported in part by the 
DOE grant
DE-FG02-ER45495.  B. C. would also like to acknowledge the hospitality of ITP,
Santa Barbara, where some of this work was performed. B. C. would
also like to thank Jens Norskov for suggesting the incompressibility argument.

\newpage

\section*{References}

\noindent Borie, B., and Sparks, C. J., 1971, Acta Crystallogr., {\bf A27}, 198.

\noindent Chakraborty, B., 1995, Europhys. Lett., {\bf 30}, 531.

\noindent Chakraborty, B., and Xi, Z., 1992, Phys. Rev. Lett., {\bf 68}, 
2039.

\noindent Daw, M. S., Foiles, S. M., and Baskes, M. I., 1993, Mat. Sci. Reports,
{\bf 9}, 251.

\noindent Dunweg, B., and Landau, D., 1993, Phys. Rev. B, {\bf 48}, 14182.

\noindent Elder, K. R., Malis, O., Ludwig, K., Chakraborty, B., and Goldenfeld, N., 1998,
accepted in Europhysics Letters.

\noindent Frenkel, A. I., Stern, E. A., Rubshtein, A., Voronel, A., and 
Rosenberg, Yu., 1997, J. Phys. IV France, {\bf 7}, C2-1005.

\noindent Hashimoto, S., 1983, Acta Crystall., 1983, A{\bf 39}, 524. 

\noindent Jacobsen, K. W., Norskov, J. K., and Puska, M. J., 1987, Phys. Rev B 
{\bf 35}, 7423.

\noindent Jacobsen, K. W., 1988, Comments Cond. Mat. Phys., {\bf 14}, 129.

\noindent Jiang, X., Ice, G. E., Sparks, C. J., Robertson, L., and Zschack, P., 1996,
Phys. Rev. B, {\bf 54}, 3211. 

\noindent Lu, Z. W., Laks, D. B., Wei, S. -H., and Zunger, A., 1994, 
Phys. Rev. B, {\bf 50}, 6642.

\noindent Malis, O., Ludwig, K.F., Schweika, W., Ice, G. E., and Sparks, C. J., 1998,
in preparation.

\noindent Metcalfe, E., and Leake, J. A., 1975, Acta Metall., {\bf 23}, 1135.

\noindent Ozolins, V., Wolverton, C., and Zunger, A., 1997, cond-mat no. 9710225.

\noindent Polatoglou, H. M., and Bleris, G. L., 1994, Solid State Commun., {\bf 
90}, 425.

\noindent Silverman, A., Zunger, A., Kalish, R., and Adler, J., 1995,  
Phys. Rev. B {\bf 51}, 10795.

\noindent Stoltze, P., 1997, Simulation methods in atomic-scale materials physics, 
(Polyteknisk Forlag, Denmark).

\noindent Xi, Z., Chakraborty, B., Jacobsen, K. W., and Norskov, J. K., 1992, 
J. Phys.: Condens. Matter, {\bf 4}, 7191.

\noindent Wolverton, C., and Zunger, A., 1995, Phys. Rev. Lett., {\bf 75}, 3162.



%
%

%
%

\end{document}